\documentclass[twocolumn,aps,pra,showpacs,floatfix]{revtex4-1}

\usepackage{amsmath}
\usepackage{amsfonts}
\usepackage{amssymb}
\usepackage{bm}
\usepackage{bbm}
\usepackage{graphicx}
\usepackage{color}
\usepackage{cancel}
\usepackage{maybemath}
\usepackage{fancybox}

\usepackage{calligra}

\newcommand{\dd}{\mathrm{d}}
\newcommand{\ee}{\mathrm{e}}
\newcommand{\ii}{\mathrm{i}}

\newcommand{\calO}{\mathcal{O}}

\newcommand{\wtilde}[1]{\mbox{$\widetilde #1$}}

\DeclareMathAlphabet{\mathcalligra}{T1}{calligra}{m}{n}
\DeclareFontShape{T1}{calligra}{m}{n}{<->s*[2.4]callig15}{}

\newcommand{\sr}{\rho}
\newcommand{\rhosymb}{\mathcal{R}}

\newcommand{\cs}{\overline\gamma}
\newcommand{\fs}{\widetilde\gamma}

\newcommand{\F}{F}

\newcommand{\RN}{{\rm RN}}
\newcommand{\FW}{{\rm (FW)}}
\newcommand{\s}{{\rm s}}
\newcommand{\Q}{{Q}}
\newcommand{\q}{{q}}

\newcommand{\rr}{\frac{r_\s}{r}}
\newcommand{\rc}{\frac{r_\Q^2}{r^2}}
\newcommand{\rrt}{\frac{r_\s}{2r}}
\newcommand{\rct}{\frac{r_\Q^2}{2r^2}}
\newcommand{\rcf}{\frac{r_\Q^2}{4r^2}}

\newcommand{\pp}{\vec p\,^2}
\newcommand{\pppp}{\vec p\,^4}

\newcommand{\alphaeff}{\alpha_{\rm eff}}

\newcommand{\adp}{\vec\alpha\cdot\vec p}
\newcommand{\sdl}{\vec\Sigma\cdot\vec L}
\newcommand{\ssdl}{\vec\sigma\cdot\vec L}

\begin{document}

\title{Dirac Hamiltonian and Reissner--Nordstr\"{o}m Metric:\\
Coulomb Interaction in Curved Space--Time}

% ===================================================================
% Dirac Hamiltonian and Reissner--Nordstrom Metric:
% Coulomb Interaction in Curved Space--Time
% ===================================================================
% J. H. Noble and U. D. Jentschura 
% ===================================================================
\author{J. H. Noble}
\affiliation{Department of Physics,
Missouri University of Science and Technology,
Rolla, Missouri 65409, USA}
% \email{jhnwdf@mst.edu}

\author{U. D. Jentschura}
\affiliation{Department of Physics,
Missouri University of Science and Technology,
Rolla, Missouri 65409, USA}

\begin{abstract} 
We investigate the 
spin-$1/2$ relativistic quantum dynamics in the curved space-time 
generated by a central massive charged object (black hole).
This necessitates a study of the coupling of a Dirac particle
to the Reissner--Nordstr\"{o}m space-time geometry and the 
simultaneous covariant coupling to the central electrostatic field.
The relativistic Dirac Hamiltonian 
for the Reissner--Nordstr\"{o}m geometry is derived.
A Foldy--Wouthuysen transformation reveals the 
presence of gravitational, and electro-gravitational
spin-orbit coupling terms which generalize the
Fokker precession terms found for the Dirac--Schwarzschild
Hamiltonian, and other electro-gravitational correction terms
to the potential proportional to $\alpha^n \, G$,
where $\alpha$ is the fine-structure constant,
and $G$ is the gravitational coupling constant.
The particle-antiparticle symmetry found for the 
Dirac--Schwarzschild geometry (and for other geometries 
which do not include electromagnetic interactions) 
is shown to be explicitly broken due to the electrostatic coupling.
The resulting spectrum of radially symmetric, 
electrostatically bound systems (with gravitational corrections)
is evaluated for example cases.
\end{abstract}


\maketitle

%
% Introduction
%
\section{Introduction}
\label{sec1}

We continue a series of 
investigations~\cite{Je2013,JeNo2013pra,Je2014pra,JeNo2014jpa,NoJe2015tach}
on the coupling of Dirac particles to curved space--time backgrounds.
Foundations of the formalism 
date back to the time of Brill and Wheeler~\cite{BrWh1957},
and Greiner, Soffel, M\"uller,
and Boulware~\cite{Bo1975prd,SoMuGr1977}, who established 
the formalism of the spin connection matrices.
Modern computer algebra \cite{Wo1999}
makes it possible to perform independent evaluations of
spin connection matrices for specific space--time 
geometries, and the formalism of the Foldy--Wouthuysen transformation 
facilitates the identification of the nonrelativistic limit,
and leads to a consistent interpretation of the
nonrelativistic operators~\cite{FoWu1950,BjDr1964}.

Recently, it has been recalled~\cite{Je2013,JeNo2013pra,Je2014pra,JeNo2014jpa,NoJe2015tach}
that the Dirac equation provides for an
ideal tool to study gravitational interactions of
antiparticles.  One should recall that the original surprising 
prediction of the Dirac equation~\cite{Di1928a,Di1928b}
was the existence of positrons,
which are the antiparticles of electrons.
The Dirac equation describes particles and their
antiparticles simultaneously.
The mass term in the Dirac equation is first and foremost the inertial
mass. However, when coupling the particle to 
curved space-time and identifying the Hamiltonian,
one can establish a connection of the inertial 
mass to the gravitational mass because the 
Foldy--Wouthuysen transformed Hamiltonian~\cite{JeNo2013pra}
contains the gravitational potential (plus relativistic corrections,
of course). On the basis of this consideration,
a symmetry relation was found in Ref.~\cite{Je2013} and 
confirmed in Ref.~\cite{JeNo2013pra} which established that
particles and antiparticles behave identically in the
presence of a gravitational field, i.e., both 
particles and antiparticles are attracted by gravity.

The symmetry relation from Refs.~\cite{Je2013,JeNo2013pra} holds for specific
classes of metrics. On the one hand, one can show that the
Reissner--Nordstr\"om metric (which describes a charged gravitational center)
belongs to a class of geometries, where  a priori, particle--antiparticle
symmetry should exist~\cite{Je2013,JeNo2013pra}, and both particles and
antiparticles should be affected identically by the metric.  On the other hand,
we know that the Dirac equation can be used 
to describe charged spin--$1/2$
particles, and that particle-antiparticle symmetry does not hold for
electromagnetic interactions~\cite{BjDr1964}. 
By definition, antiparticles carry the opposite
electric charge.  How can this apparent contradiction be resolved?
The answer is that the presence of the explicit covariant coupling to the 
electrostatic field, not to the gravitational
field, breaks the particle-antiparticle symmetry. 
It means that we must concern ourselves with 
the coupling of the Dirac particle to the
curved space--time,
while at the same time include the electrostatic interaction.
The Dirac equation becomes covariant with respect to two gauge 
groups, the $U(1)$ gauge group of quantum electrodynamics
and the $SO(1;3)$ group of local Lorentz transformations.
The double-covariant derivative entails the replacement
$\ii \partial_\mu \to \ii (\partial_\mu - \Gamma_\mu) - q \, A_\mu$,
where $\Gamma_\mu$ is the spin-connection matrix,
$q$ is the charge of the particle, and $A_\mu$ is the vector potential~\cite{No2015phd}.
The former covariance is ensured by the four-vector 
potential $A^\mu$ in the Dirac equation,
while the latter is described by the spin-connection matrices
$\Gamma_\mu$, both to be discussed in more detail below.

Bound systems featuring both electromagnetic as well 
as gravitational corrections could be of interest 
for a number of reasons, not only in the sense
of tiny gravitational effects which might be observable
in bound systems~\cite{ChSJTu2001},
but also in the context of micro black holes which 
have been proposed as conceivable 
candidates for dark matter~\cite{GrBaAl2004,GrBa2008,DoEr2014}
and even classes of novel phenomena at
accelerators~\cite{JaBuWiSa2000}.

The article is organized as follows.
In Sec.~\ref{sec2}, we transform the Reissner--Nordstr\"om metric
into isotropic coordinates, and transform the electrostatic
potential accordingly.
These results are then used in Sec.~\ref{sec3a} in the explicit
derivation of the Dirac--Reissner--Nordstr\"om Hamiltonian.
We then apply the Foldy--Wouthuysen transform to the
resulting Hamiltonian in Sec.~\ref{sec3b}.
In Sec.~\ref{sec4}, we evaluate the bound--state energies of
the transformed Hamiltonian, and consider example cases.
Finally, conclusions are drawn in Sec.~\ref{sec5}.
Except where otherwise stated,
we use units such that $c=\hbar=\epsilon_0=1$ throughout this paper.

%
% Reissner--Nordstr\"{o}m Metric and Electrostatic Potential
%
\section{Reissner--Nordstr\"{o}m Metric and Electrostatic Potential}
\label{sec2}

%
% Isotropic Coordinates
%
\subsection{Isotropic Coordinates}
\label{sec2a}

In formulating the Dirac equation coupled to the Reissner--Nordstr\"om metric
we can in large part follow the same steps taken to formulate the
gravitationally coupled Dirac Hamiltonian~\cite{JeNo2013pra}.
This also provides the opportunity to check our final result against results 
previously obtained, when the charge of the gravitational center vanishes
($Q\to0$) and the transformed Dirac--Reissner--Nordstr\"om Hamiltonian 
reduces to the transformed Dirac--Schwarzschild Hamiltonian found in 
Eq.~(21) of Ref.~\cite{JeNo2013pra}.
We require that the metric be isotropic,
in order to ensure that the effective speed of light,
expressed in global coordinates,
does not depend on the spatial direction of the light ray
at a given space-time point
(note that the speed of light is not constant when 
expressed in global coordinates, a fact 
which in particular, allows for the existence of the 
Shapiro time delay~\cite{Sh1964,ShEtAl1968,Sh1999,Lo1988}).
We follow ideas of Eddington~\cite{Ed1924}
and transform the Reissner--Nordstr\"om metric 
to isotropic coordinates.
The derivation of the Reissner--Nordstr\"om metric 
is recalled in Appendix~\ref{appa}, with the result
\begin{align}
\dd s^2=&\,\left(1-\frac{r_\s}{\rhosymb}+\frac{r_\Q^2}{\rhosymb^2}\right)\dd t^2
\nonumber\\&\,
-\left(1-\frac{r_\s}{\rhosymb}+\frac{r_\Q^2}{\rhosymb^2}\right)^{-1}\dd\rhosymb^2
-\rhosymb^2\,\dd\Omega^2\,,
\label{RNmetric}
\end{align}
where $r_\s=2\,G\,M/c^2$ is the Schwarzschild radius and 
$r_\Q^2=G\,Q^2/ (4 \pi \epsilon_0 c^4)$
(we temporarily restore SI mksA units for the conversions).
In order to convert the Reissner--Nordstr\"om metric into a spatially isotropic
form, we use the transformation
\begin{equation}
\label{IsoTrans}
\rhosymb=r\left(\left(1+\frac{r_\s}{4r}\right)^2-\frac{r_\Q^2}{4r^2}\right)
=r\,A(r)\,.
\end{equation}
Under this transform, we find
\begin{equation}
\dd\rhosymb=\left(\left(1-\frac{r_\s}{4r}\right)\left(1+\frac{r_\s}{4r}\right)
+\frac{r_\Q^2}{4r^2}\right)\dd r=B(r)\,\dd r\,,
\end{equation}
and
\begin{equation}
1-\frac{r_\s}{\rhosymb}+\frac{r_\Q^2}{\rhosymb^2}
=\frac{\left(\left(1-\frac{r_\s}{4r}\right)\left(1+\frac{r_\s}{4r}\right)
+\frac{r_\Q^2}{4r^2}\right)^2}
{\left(\left(1+\frac{r_\s}{4r}\right)^2-\frac{r_\Q^2}{4r^2}\right)^2}
=\frac{B(r)^2}{A(r)^2}\,.
\end{equation}
The metric becomes
\begin{align}
\dd s^2=&\,\frac{B(r)^2}{A(r)^2}\dd t^2-\frac{A(r)^2}{B(r)^2}B(r)^2\dd r^2
-r^2\,A(r)^2\dd\Omega\nonumber\\
=&\,\frac{B(r)^2}{A(r)^2}\dd t^2-A(r)^2\left(\dd r^2+r^2\dd\Omega^2\right)\,,
\end{align}
i.e.,
\begin{subequations}
\label{isoRNmetric}
\begin{align}
\label{wv1}
\dd s^2 =& \,w(r)^2\dd t^2-v(r)^2\left(\dd x^2+\dd y^2+\dd z^2\right)\,,\\
\label{wv2}
w(r)=&\,\frac{\left(1-\frac{r_\s}{4r}\right)\left(1+\frac{r_\s}{4r}\right)
+\frac{r_\Q^2}{4r^2}}
{\left(1+\frac{r_\s}{4r}\right)^2-\frac{r_\Q^2}{4r^2}}\,,\\
\label{wv3}
v(r)=&\,\left( 1+\frac{r_\s}{4r}\right)^2-\frac{r_\Q^2}{4r^2}
\end{align}
\end{subequations}
As in Refs.~\cite{JeNo2013pra,JeNo2014jpa,NoJe2015tach},
we keep terms only to the first order in $G$.
Both $r_\s$ and $r_\Q^2$ are proportional to $G$;
hence, $w(r)$ and $v(r)$ are approximated to
\begin{align}
\label{approxwv}
w(r)\approx1-\rrt+\rct\,,\quad
v(r)\approx1+\rrt- \frac{r_\Q^2}{4 r^2}\,.
\end{align}
In the limit $Q\to0$, we recover the $w(r)$ and $v(r)$ from the Schwarzschild metric
[see Eq.~(14) of Ref.~\cite{JeNo2013pra}].

\subsection{Electrostatic Potential}
\label{sec2b}

As shown in Appendix~\ref{appa}, 
the nonzero elements of the field strength tensor are
\begin{equation}
F_{t\rhosymb}=-F_{\rhosymb t}=\frac{Q}{4\pi\,\rhosymb^2}\,.
\end{equation}
By definition [see Eq.~(2.2.28) of Ref.~\cite{Wa1984}]
the field strength tensor is given as
\begin{equation}
F_{\mu\nu}=\partial_\mu A_\nu-\partial_\nu A_\mu\,.
\end{equation}
We then find that the resulting equation is solved by
\begin{equation}
A_0=\frac Q{4\pi\,\rhosymb}\,,\quad \vec A=\vec0\,.
\end{equation}
Applying the isotropic transform [Eq.~\eqref{IsoTrans}] 
to our potential we obtain
\begin{equation}
\label{A0}
A_0=\frac{Q}{4\pi\,r\left(\left(1+\frac{r_\s}{4r}\right)^2
-\frac{r_\Q^2}{4r^2}\right)}\,.
\end{equation}
Again, we are keeping terms only to the first order in $G$.
Thus, when expressed in terms of the isotropic 
radial coordinate $r$, we have
\begin{equation}
\label{IsoPot}
A_0=\frac Q{4\pi\,r}\left(1-\rrt+\rcf\right)
\end{equation}
for the electrostatic potential.

%
% Dirac Hamiltonian for the Reissner--Nordstr\"{o}m Metric
%
\section{Dirac Hamiltonian for the Reissner--Nordstr\"{o}m Metric}
\label{sec3}

%
% Relativistic Hamiltonian
%
\subsection{Relativistic Hamiltonian}
\label{sec3a}

In order to derive the Dirac Hamiltonian for the 
Reissner--Nordstr\"{o}m metric,
one uses the double-covariant coupling prescription
$\ii \partial_\mu \to \ii (\partial_\mu - \Gamma_\mu) - q \, A_\mu$
where $\Gamma_\mu$ is the spin-connection 
matrix and $A_\mu$ is the electrostatic potential,
both expressed in isotropic coordinates.
Using the form given in
Eqs.~\eqref{wv1} and~\eqref{approxwv} for the Reissner--Nordstr\"{o}m metric,
one readily evaluates the spin-connection matrices $\Gamma_\mu$ 
using general formulas
and inserts $A_\mu$ from Eq.~\eqref{IsoPot}.
The technical details of the calculation can be 
found in Appendix~\ref{appb}.
The Dirac--Reissner--Nordstr\"om Hamiltonian, to the first order 
in $G$, is finally found as 
\begin{align}
H_\RN =&\,\frac12\left\{\adp,\left(1-\rr+
\frac{3 r_\Q^2}{4 r^2} \right)\right\}
\nonumber\\&\,
+\frac{Z_\Q\, Z_\q\,\alpha}r\left(1-\rrt+\rcf\right)
\nonumber\\&\,
+\beta\,m\left(1-\rrt+\rct\right)\,,
\end{align}
where in natural units, we have
\begin{equation}
q\,Q=4\pi\,Z_\Q\,Z_\q\,\alpha \,.
\end{equation}
$Z_\Q$ and $Z_\q$ are the nuclear charge numbers associated with $Q$
and $q$, respectively. For $Q\to0$ (which implies $Z_\Q\to0$ and $r_\Q\to0$),
we recover the Dirac--Schwarzschild Hamiltonian~\cite{JeNo2013pra}.

% Moving forward we return to units where $c=\hbar=\epsilon_0=1$.

%
% Foldy--Wouthuysen Transformation
%
\subsection{Foldy--Wouthuysen Transformation}
\label{sec3b}

An exact Foldy--Wouthuysen transformation may be used 
in the case of the free Dirac Hamiltonian \cite{BjDr1964}.
More complicated
Hamiltonians require a perturbative approach, expanding 
in terms of some perturbation parameter.
Using an approach 
similar to the steps taken in Eq.~(3) of 
Ref.~\cite{Je2013}, we define the
dimensionless variable $\sr$ in terms of the fine structure constant, as
\begin{equation}
\label{defalphaeff}
\sr= \frac{r}{a_0} \, \quad
a_0 = \frac{\hbar}{\alphaeff\,m c} \,,\quad
\alphaeff=\frac{q\,Q}{4\pi\,\epsilon_0\,\hbar\,c}
=Z_\Q\,Z_\q\,\alpha\,,
\end{equation}
where we have temporarily implemented SI mkSA units for the sake of clarity
($a_0$ is a generalized Bohr radius, while $\alphaeff$ is an effective
``fine-structure'' constant, i.e., coupling constant,
for the bound system of charged black 
hole and test particle). Then, in natural units,
\begin{subequations}
\label{rpalpha}
\begin{align}
\label{rpalphaa}
r=&\,\frac 1{\alphaeff\,m}\sr\,,\quad
\vec\nabla_r=\alphaeff\,m\,\vec\nabla_\sr\,,\\
\label{rpalphab}
\vec p=&\,-\ii\,\alphaeff\,m\,\vec\nabla_\sr\,.
\end{align}
\end{subequations}
We then use $\alphaeff$ as our expansion parameter in our calculation,
keeping terms up to $\alphaeff^4$, and to the first order in the gravitational
interaction ($G$), i.e., we keep
all terms up to order $\alphaeff^4$, and $\alphaeff^4 \, G$.
E.g., momentum operators contribute 
one power of $\alphaeff$, according to Eq.~\eqref{rpalphab}.
The parameter $r_\Q^2$, where
\begin{equation}
r_\Q^2 = \frac{G\,Q^2}{4 \pi} = G \, Z_\Q^2 \alpha \,,
\end{equation}
is counted as a single power of $G$, because $Z_\Q^2$ may be large,
resulting in $Z_\Q^2 \alpha$ being of order unity.
Terms of second order in the gravitational interaction ($G^2$) are
ignored. This is consistent with the approximations made earlier
in this article, namely, in 
Eqs.~\eqref{approxwv} and~\eqref{IsoPot}.

In applying the Foldy--Wouthuysen transformation, we first identify
the odd part (in bispinor space) of the Hamiltonian $H_\RN$
\begin{equation}
\calO=\frac12\left\{\adp,\left(1-\rr+ \frac{3 r_\Q^2}{4 r^2} \right)\right\}\,.
\end{equation}
We now construct the Hermitian operator $S$ and the unitary transform $U$ as
\begin{equation}
S=-\ii\frac{\beta\,\calO}{2m}\,,\quad
U=\exp\left(\ii\,S\right)\,.
\end{equation}
We can now apply the first iteration of
the Foldy Wouthuysen transform using the approximation
\begin{align}
H'=&\,U\,H_\RN\,U^{\rm +}
=\ee^{\ii\,S}\,H_\RN\,\ee^{-\ii\,S}
\nonumber\\
=&\,H_\RN+\ii\left[S,H_\RN\right]
+\frac{\ii^2}{2!}\left[S,\left[S,H_\RN\right]\right]
%+\frac{\ii^3}{3!}\left[S,\left[S,\left[S,H_\RN\right]\right]\right]
+\dots\,.
\end{align}
We perform the transformation and calculate
\begin{align}
\label{implicit}
H'
=&\,\beta\left(m+\frac{\calO^2}{2m}-\frac{\calO^4}{8m^3}\right)
+\frac{Z_\Q\,Z_\q\,\alpha}r\left(1-\rrt+\rcf\right)
\nonumber\\&\,
-\frac1{8m^2}\left[\calO,\left[\calO,\frac{Z_\Q\,Z_\q\,\alpha}r\right]\right]
-\beta\frac{m\,r_\s}{2r}
+\beta\frac{m\,r_\Q^2}{2r^2}
\nonumber\\&\,
+\frac{\beta}{16m}\left\{\calO,\left\{\calO,\rr-\rc\right\}\right\}
+\calO'\,,
\end{align}
where
\begin{align}
\calO'
=&\,-\frac{\calO^3}{3m^2}
+\frac{\beta}{2m}\left[\calO,\frac{Z_\Q\,Z_\q\,\alpha}
r\left(1-\rrt\right)\right]
\nonumber\\&\,
+\frac14\left\{\calO,\rr-\rc\right\}
-\frac1{96m^2}\left\{\calO,\left\{\calO,\left\{\calO,
\rr\right\}\right\}\right\}\,.
\end{align}
Notice that the leading-order terms in $\calO$ are of order
$\alphaeff$ while the leading-order terms in $\calO'$
are of order $\alphaeff^3$ and $\alphaeff^2 \, G$.
In Eq.~\eqref{implicit}, we have several multi-commutators 
involving $\calO$.
The implicit understanding is that terms of higher order than 
$\alpha_{\rm eff}^4$ and $\alpha_{\rm eff}^4 \, G$ 
generated by these multi-commutators can be neglected.
Each iteration of the Foldy--Wouthuysen transform eliminates terms
up to the leading order of the odd part, but may introduce higher order
odd terms. 
In iterating the procedure the odd terms are eventually eliminated up to
a desired order.
Applying the transform to $H'$ will give us the Hamiltonian $H''$ with
odd part $\calO''\sim\alphaeff^4G$.
One further iteration will fully eliminate the odd terms up to order
$\alphaeff^4$ and first order in $G$.
For an iteration to contribute to the even part of the Hamiltonian
the square of the odd part associated with that iteration must be
within the desired order. 
Because the leading terms in $\calO'^2$
are of order $\alphaeff^5G$ and $\alphaeff^6$,
they can be ignored within our approximations.
Thus our Foldy--Wouthuysen transform of the Dirac--Reissner--Nordstr\"om
Hamiltonian requires three iterations.
The first of these determines the form of the even part, while
the final two serve to fully eliminate the odd part (up to our desired order).

The three-fold iterated
Foldy--Wouthuysen transformation then gives us
\begin{align}
H_\RN^\FW
=&\,\beta\left(m+\frac{\calO^2}{2m}-\frac{\calO^4}{8m^3}\right)
\nonumber\\&\,
+\frac{Z_\Q\,Z_\q\,\alpha}r\left(1-\rrt+\rcf\right)
\nonumber\\&\,
-\frac1{8m^2}\left[\calO,\left[\calO,\frac{Z_\Q\,Z_\q\,\alpha}r\right]\right]
-\beta\frac{m\,r_\s}{2r}
+\beta\frac{m\,r_\Q^2}{2r^2}
\nonumber\\&\,
+\frac{\beta}{16m}\left\{\calO,\left\{\calO,\rr-\rc\right\}\right\}\,.
\end{align}
Finally, we calculate all the terms involving the original odd part $\calO$,
giving the final result
\begin{align}
\label{HFWRN}
H_\RN^\FW
=&\,\beta\left(m+\frac\pp{2m}-\frac\pppp{8m^3}\right)
\nonumber\\&\,
+\frac{Z_\Q\,Z_\q\,\alpha}r\left(1-\rrt+\rcf\right)
\nonumber\\&\,
-\frac{Z_\Q\,Z_\q\,\alpha\,\pi}{2m^2}\delta^{(3)}\left(\vec r\right)
-\frac{Z_\Q\,Z_\q\,\alpha}{4m^2}\frac\sdl{r^3}
\nonumber\\&\,
-\beta\frac m2\left(\rr-\rc\right)
-\beta\frac 3{8m}\left\{\pp,\rr- \frac{2 r_Q^2}{3 r^2} \right\}
\nonumber\\&\,
+\beta\frac{3\pi\,r_\s}{4m}\delta^{(3)}\left(\vec r\right)
+\beta\frac{3}{8m}\frac\sdl{r^2}\left(\rr-
\frac{4 r_Q^2}{3 r^2} \right)
\nonumber\\&\,
+\beta \, \frac{r_\Q^2}{4 \, m\,r^4}
-\beta\frac{\pi\,r_\Q^2}{m\,r}\delta^{(3)}(\vec r) \,.
\end{align}
A few remarks are in order. For a reference $S$ state,
the expectation values of the operators 
$\{ \vec p^{\,2}, 1/r^2 \}$,
$1/r^4$, and $ \delta^{(3)}(\vec r) /r$ diverge. 
In this article, we shall explicitly exclude $S$ states from the analysis 
and concentrate on highly excited Rydberg states for which 
the expectation value of $\delta^{(3)}(\vec r)/r$  vanishes~\cite{Je2014pra}.
The emergence of this operator is a consequence of the point nucleus 
approximation inherent to the Coulomb potential,
which is manifest in the divergence of the scalar potential $A_0$ 
given in Eq.~\eqref{A0} for $r \to 0$. For a realistic nucleus (a realistic central charged 
black hole), this divergence is cut off due to the nuclear 
finite-size effect, and the operator $ \delta^{(3)}(\vec r) /r$ would 
need to be replaced by a term proportional to $V_n(r) \vec\nabla^{\,2} V_n(r)$,
where $V_n(r)$ is the nuclear potential including the finite-size effect~\cite{MoSo1993,SoMo1988}.
We now return to the analysis of the result given in Eq.~\eqref{HFWRN}.
For $Q\to0$ ($Z_\Q\to0$),
we recover the Foldy--Wouthuysen transformed 
Dirac--Schwarzschild Hamiltonian found in Eq.~(21) of Ref.~\cite{JeNo2013pra}.
Alternatively, we can rewrite the transformed Hamiltonian as
\begin{align}
H_\RN^\FW=&\,H^\FW_{\rm F}+H'_{\rm DC}+H'_{\rm DS}\,,
\end{align}
where $H^\FW_\F$ is the free Hamiltonian,
\begin{equation}
\label{HF}
H^\FW_\F =
\beta\left(m+\frac\pp{2m}-\frac\pppp{8m^3}\right) \,.
\end{equation}
$H'_{\rm DC}$ is a gravitationally 
modified Dirac--Coulomb Hamiltonian without the kinetic terms
which are summarized in $H^\FW_\F$,
\begin{align}
\label{HDCp}
H'_{\rm DC}=&\, \frac{Z_\Q\,Z_\q\,\alpha}r\left(1-\rrt+\rcf\right)
\nonumber\\&\,
-\frac{Z_\Q\,Z_\q\,\alpha\pi}{2m^2}\delta^{(3)}(\vec r)
-\frac{Z_\Q\,Z_\q\,\alpha}{4m^2}\frac{\sdl}{r^3}\,,
\end{align}
Moreover, $H'_{\rm DS}$ is an electromagnetically
modified Dirac--Schwarzschild Hamiltonian,
again without the kinetic terms
which are found in $H^\FW_\F$,
\begin{align}
\label{HDSp}
H'_{\rm DS}=&\,-\beta\frac m2\left(\rr-\rc\right)
-\beta\frac{3}{8m}\left\{\pp,\rr- \frac{2 r_Q^2}{3 r^2}\right\}
\nonumber\\&\,
+\beta\frac{3\pi\,r_\s}{4m}\delta^{(3)}(\vec r)
+\beta\frac{3}{8m}\frac{\sdl}{r^2}\left(\rr-
\frac{4 r_Q^2}{3 r^2} \right)
\nonumber\\&\,
+\beta \frac{r_\Q^2}{4 \, m\,r^4}
-\beta\frac{\pi\,r_\Q^2}{m\,r}\delta^{(3)}(\vec r)\,.
\end{align}
Up to the electromagnetic modifications 
of the gravitational terms in the Dirac--Schwarzschild
Hamiltonian, and up to the 
gravitational modifications of the 
Dirac--Coulomb Hamiltonian, we thus have
$H'_{\rm DC}\approx H_{\rm DC}^\FW-H_\F^\FW$ and
$H'_{\rm DS}\approx H_{\rm DS}^\FW-H_\F^\FW$,
where $H_{\rm DC}^\FW$ and 
$H_{\rm DS}^\FW$ are given in Eqs.~(30) and (47) 
of Ref.~\cite{JeNo2014jpa}. 
The relativistic corrections found in the transformed Reissner--Nordstr\"om 
Hamiltonian are approximately equal to a 
sum of the corrections found for 
the Dirac--Coulomb and the Dirac--Schwarzschild Hamiltonians,
with additional electro-gravitational mixing terms
(the latter are proportional to the product of the 
gravitational coupling constant, and a power of the 
fine-structure constant). Both $H_\F^\FW$ and
$H'_{\rm DS}$ exhibit particle--antiparticle symmetry
(all terms have a $\beta$ prefactor),
while $H'_{\rm DC}$ changes sign 
under particle--antiparticle interchange
(no $\beta$ prefactor).

\begin{figure}[t!]
\centerline{\includegraphics[width=0.7\linewidth]{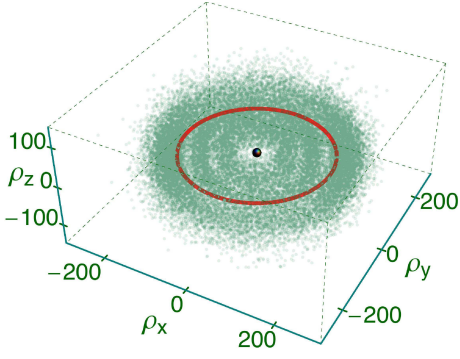}}
\caption{\label{fig1}
(Color online.)
Scatter plot of the probability density of finding 
a bound particle (electron) in a state of $n = 12$,
$\ell = 9$ and $m = |\ell| = 9$ in the field of 
a charged heavy black hole
with mass $10^{-12}$ times the 
mass of the earth,
and charge number $Z_\Q = 10$.
The points are distributed randomly,
with the number of scattered points in 
a reference volume being proportional to the 
probability of finding the bound electron
in the volume.
The two radial minima of the probability density 
are clearly visible. 
The gravitational center is depicted as a black dot.
For reference, the classical trajectory 
at $\langle \rho \rangle = \int \rho \, | \psi(\vec \rho) |^2 \; \dd^3 r = 
171$ is also shown.
Note that the scaled radial coordinate $\vec\rho$ given in Eq.~\eqref{scaling}
is dimensionless, as reflected in the labeling of the axes.}
\end{figure}

%
% Bound--State Energies
%
\section{Bound--State Energies}
\label{sec4}

It remains to evaluate and discuss the bound-state energies in the
potential described by Eq.~\eqref{HFWRN},
and to consider an example case.
First, we observe that the product $Z_\Q \, Z_\q$ has to be negative for the 
electrostatic interaction to be attractive and bound states
to exist. We thus define the coupling constants,
\begin{subequations}
\begin{align}
\alphaeff =& \; - Z_\Q \, Z_\q \, \alpha > 0 \,,
\\
\label{defalphaG}
\alpha_G =& \; \frac{G m M}{\hbar c} = G m M \,,
\\
\label{defalphaQ}
\alpha_Q =& \; r_\Q^2 \left( \frac{m c}{\hbar} \right)^2
= \frac{Z_\Q^2 \, e^2\, G m^2}{4 \pi \epsilon_0 \hbar^2 c^2}
= Z_\Q^2 \,\alpha \, G m^2 \,,
\end{align}
\end{subequations}
where in the intermediate steps we 
temporarily restore full SI mksA units.
Following Ref.~\cite{Je2014pra}, it is advantageous
to scale the coordinate variable 
according to 
\begin{equation}
\label{scaling}
\vec \rho = \alphaeff \, m \, \vec r \,,
\qquad
\vec \nabla \equiv
\vec \nabla_r = \alphaeff \, m \, \vec \nabla_\rho \,,
\end{equation}
where $\vec \rho$ is the coordinate in ``atomic units'';
the ``Bohr radius'' is $( \alphaeff \, m )^{-1}$.

\begin{widetext}

From Eq.~\eqref{HFWRN}, we first extract the
effective Hamiltonian applicable to particle 
(as opposed to antiparticle) states
[hence denoted with a superscript $(+)$],
and scale the expression according to Eq.~\eqref{scaling},
\begin{align}
H_\RN^{(+)}
=&\,m+\alphaeff^2\,m \, \left(-\frac12\vec\nabla_\rho^2-\frac1\rho\right)
+ \alphaeff^4\,m\left(-\frac18\vec\nabla^4_\rho
+\frac{\pi}{2}\,\delta^{(3)}(\vec \rho)
+\frac{\ssdl}{4 \rho^3}\right)
- \frac{\alpha_G\,\alphaeff\,m}{\rho}
+ \alpha_G\,\alphaeff^3\,m\left(
\frac34\left\{\vec\nabla_\rho^2,\frac1\rho\right\}
\right.
\nonumber\\
&\, 
\left.
+\frac{3 \pi}2\delta^{(3)}(\vec\rho)
+\frac{3 \, \ssdl}{4 \rho^3} 
+ \frac{1}{\rho^2} \right)
+\frac{\alpha_Q\,\alphaeff^2\,m}{2 \rho^2}
+ \alpha_Q\,\alphaeff^4\,m\left(
-\frac1{4 \rho^3}
-\frac14 \left\{\vec\nabla_\rho^2,\frac1{\rho^2}\right\}
-\frac{\ssdl}{2 \, \rho^4}
+\frac{1}{4\,\rho^4}-\frac\pi{8\rho}\delta^{(3)}(\vec\rho) \right) \,.
\end{align}
Let us break down the matrix elements for the energy corrections
in an unperturbed Dirac--Coulomb state with 
quantum numbers $n$, $\ell$ and $j$
according to the Hamiltonians in Eqs.~\eqref{HF},~\eqref{HDCp} and~\eqref{HDSp}.
An evaluation using formulas given in Ref.~\cite{BeSa1957} leads to 
the results
\begin{subequations}
\begin{align}
\langle H_\F^{(+)} \rangle =& \;
m \, \left\{ 1 + \frac{\alphaeff^2}{2 n^2} + 
\alphaeff^4  \, \left( \frac{3}{8 n^4} - \frac{1}{n^3 (2\ell + 1)}
\right) \right\} \,,
\\
\langle {H'}_{\rm DC}^{(+)} \rangle =& \;
m \, \left\{ - \frac{\alphaeff^2}{n^2} +
\alphaeff^4 \left( 
- \frac{\delta_{j, \ell - 1/2}}{2 n^3 \ell (2\ell +1)} + 
\frac{\delta_{j, \ell + 1/2}}{2 n^3 \ell (\ell + 1) \, (2\ell +1)}
\right)
+ \frac{2 \alpha_G \, \alphaeff^3}{n^3 \, (2 \ell + 1)}
- \frac{\alpha_Q \, \alphaeff^4 }{2 \, n^3 \, \ell \, (\ell + 1) \, (2 \ell + 1)}
\right\} \,,
\\
\langle {H'}_{\rm DS}^{(+)} \rangle =& \;
m \, \left\{ \alpha_G 
\left(
- \frac{\alphaeff}{n^2} 
+ \frac{\alphaeff^3}{n^2} 
\left[ \delta_{j, \ell - 1/2} \,
\left( \frac{3}{2 n^4} - \frac{3 \, (4 \ell + 1)}{2 n^3 \ell (2\ell + 1)}
\right)
+ \delta_{j, \ell + 1/2} \,
\left( \frac{3}{2 n^4} - \frac{3 \, (4 \ell + 3)}{2 n^3 (\ell + 1) (2\ell + 1)}
\right)
\right]
\right)
\right.
\nonumber\\
& \, 
+ \alpha_Q \,
\left(
\frac{\alphaeff^2}{ n^3 \, ( 2 \ell + 1 )} 
+ \alphaeff^4
\left[ \delta_{j, \ell - 1/2} \,
\left( -\frac{2 \ell }%
{n^5 (2 \ell -1 )\, (2 \ell + 1)} +
\frac{4 \ell + 1}%
{n^3 \, \ell \, (\ell + 1) \, (2 \ell -1 )\, (2 \ell + 1)} 
\right)
\right.
\right.
\nonumber\\
& \, + \left. \left. \left.
\delta_{j, \ell + 1/2} \,
\left( -\frac{2 (\ell + 1) }%
{n^5 (2 \ell - 1)\, (2 \ell + 1)} +
\frac{4 \ell + 3}%
{n^3 \, \ell \, (\ell+1) \, (2 \ell - 1)\, (2 \ell + 1) } 
\right)
\right]
\right)
\right\} \,,
\end{align}
\end{subequations}
where the functional form is seen to depend on the 
relative orientation of orbital angular momentum and spin.
Formulas become more compact when expressed in terms 
of the Dirac angular quantum number $\varkappa = 
(-1)^{\ell + j + 1/2} \, (j + \tfrac12)$,
which is defined as the negative of the eigenvalue of the 
Dirac angular operator $K = \beta \, (\vec \Sigma \cdot \vec L + 1)$,
i.e., $K \, \psi = -\varkappa \, \psi$
(see Refs.~\cite{SwDr1991a,*SwDr1991b,*SwDr1991c} for 
further discussion).
The results read as follows,
\begin{subequations}
\begin{align}
\langle H_\F^{(+)} \rangle =& \;
m \, \left\{ 1 + \frac{\alphaeff^2}{2 n^2} + 
\alphaeff^4  \, \left( \frac{3}{8 n^4} - \frac{\varkappa}{|\varkappa| \, n^3 (2\varkappa+ 1)}
\right) \right\} \,,
\\
\langle {H'}_{\rm DC}^{(+)} \rangle =& \;
m \, \left\{ - \frac{\alphaeff^2}{n^2} -
\frac{\alphaeff^4}{2 |\varkappa| n^3 \, (2 \varkappa + 1)}
+ \frac{2 \alpha_G \, \alphaeff^3 \, \varkappa}{|\varkappa| \, n^3 \, (2 \varkappa + 1)}
- \frac{\alpha_Q \, \alphaeff^4 }{2 \, |\varkappa| \, n^3 \, (\varkappa + 1) \, (2 \varkappa + 1)}
\right\} \,,
\\
\langle {H'}_{\rm DS}^{(+)} \rangle =& \;
m \, \left\{ \alpha_G
\left(
- \frac{\alphaeff}{n^2}
+ \alphaeff^3
\left[ \frac{3}{2 n^4} -
\frac{3 \, (4\varkappa + 1)}{2 |\varkappa| \, n^3 \, (2\varkappa + 1)}
\right]
\right)
+ \alpha_Q \, \left[
\frac{\alphaeff^2 \, \varkappa}{|\varkappa| \, n^3 \, (2 \varkappa + 1)} 
\right.
\right.
\nonumber\\
& \, 
\left. 
\left. 
+ \alphaeff^4 \left(
-\frac{2 \varkappa^2 }%
{|\varkappa| \, n^5 \, (2 \varkappa - 1)\, (2 \varkappa + 1)} +
\frac{\varkappa \, (4 \varkappa + 1)}%
{|\varkappa| \, n^3 \, (\varkappa + 1) \, 
(2 \varkappa -1 )\, (2 \varkappa + 1) } 
\right)
\right]
\right\} \,.
\end{align} 
\end{subequations}
An important check consists in the verification of the Dirac--Coulomb energy,
which is obtained as the sum of the $\alphaeff^4$ term from 
$\langle H_\F^{(+)} \rangle$ and the $\alphaeff^4$ term from 
$\langle {H'}_{\rm DC}^{(+)} \rangle$,
\begin{equation}
\left. \langle H_\F^{(+)} \rangle \right|_{\alphaeff^4} +
\left. \langle {H'}_{\rm DC}^{(+)} \rangle \right|_{\alphaeff^4} =
\left( \frac{3}{8 n^4} - \frac{\varkappa}{|\varkappa| \, n^3 (2\varkappa+ 1)} \right) -
\left( \frac{1}{2 |\varkappa| \, n^3 \, (2\varkappa + 1)} \right)
= \frac{3}{8 n^4} - \frac{1}{2 |\varkappa| n^3} \,.
\end{equation}
The latter result is in agreement with the literature 
[see, e.g., Eq.~(2.87) of Ref.~\cite{ItZu1980}].
The relativistic corrections to the Dirac--Schwarzschild energy
are obtained as a sum of the $\alphaeff^4$ coefficient 
of $\langle H_\F \rangle$ and the $\alpha_G \,\alphaeff^3$ term from 
$\langle H'_{\rm DS} \rangle$,
\begin{equation}
\label{confDS}
\left. \langle H_\F^{(+)} \rangle \right|_{\alphaeff^4} + 
\left. \langle {H'}_{\rm DS}^{(+)} \rangle \right|_{\alpha_G \, \alphaeff^3} =
\left( \frac{3}{8 n^4} - \frac{\varkappa}{|\varkappa| \, n^3 (2\varkappa+ 1)} \right) +
\left( \frac{3}{2 n^4} -
\frac{3 \, (4\varkappa + 1)}{2 |\varkappa| \, n^3 \, (2\varkappa + 1)} \right)
= 
\frac{15}{8 n^4} -
\frac{14 \varkappa + 3}{2 |\varkappa| \, n^3 \, (2\varkappa + 1)} \,.
\end{equation}
[We note that in the absence of the electrostatic potential,
we would scale the radial variable differently, 
namely, $\vec r = \alpha_G m \, \vec \rho$, leading to the 
sum of coefficients in the final $\alpha_G^4$ term 
as indicated in Eqs.~(3a),~(3b) and~(12) of Ref.~\cite{Je2014pra}.]
The final result given in Eq.~\eqref{confDS} is just the Dirac--Schwarzschild formula
[Eq.~(12) of Ref.~\cite{Je2014pra}].

For Rydberg states with a vanishing probability density 
at the origin (see Figs.~\ref{fig1} and~\ref{fig2}),
the influence of the black hole at the center 
can be described by a small (complex rather than real) energy correction
(see Sec.~IV of Ref.~\cite{Je2014pra}).
We have performed the scaling of the radial variable 
according to Eq.~\eqref{scaling}, implicitly assuming that the 
electrostatic interaction corresponding to the 
coupling constant $\alphaeff$ dominates over the gravitational
terms (coupling constant $\alpha_G$);
or, otherwise, we would have had to perform the 
initial scaling to the ``Bohr radius'' of the system
differently, as indicated above. One might think that, 
for the original expansion in powers of $\alphaeff$ and 
$\alpha_G$ to be valid,
$\alphaeff$ should not be too small in comparison to $\alpha_G$,
or else, before encountering any gravitational
effect, we should first include higher-order  terms in the $\alphaeff$
expansion. Fortunately, the bound-state 
theory of atoms permits us to sum all the 
relativistic corrections due to the electrostatic 
central potential into a convenient all-order 
(in $\alphaeff$) formula, which reads as 
\begin{align}
\label{ERN}
E_{\rm RN} =& \; m \left\{ f(n,\varkappa ) +
\alpha_G \, \left[ - \frac{\alphaeff}{n^2} +
\alphaeff^3 \, \left( \frac{3}{2 n^4} - 
 \frac{8 \varkappa + 3}{2 | \varkappa | \, n^3 \, (2 \varkappa + 1) }
\right) \right] 
+ \alpha_Q \, \left[ \alphaeff^2 \, 
\frac{\varkappa}{| \varkappa | \, n^3 \, (2 \varkappa + 1)}
\right.
\right.
\nonumber\\
& \; \left.  \left. 
+ \alphaeff^4 \, 
\left( - \frac{2 \varkappa^2}%
{|\varkappa| \, n^5 \, (2 \varkappa - 1)  \, (2\varkappa + 1) }
+ \frac{3}%
{2 | \varkappa| \, n^3 \, (\varkappa + 1) \, (2 \varkappa - 1)}
 \right) \right] \right\} \,,
\nonumber\\
f(n, \varkappa) =& \; 
\left( 1 + \frac{\alphaeff^2}{(n - |\varkappa| + \sqrt{\varkappa^2 - \alphaeff^2})^2}
\right)^{-1/2} =
1 - \frac{\alphaeff^2}{2 n^2} + 
\alphaeff^4 \, \left( \frac{3}{8 n^4} - \frac{1}{2 |\varkappa| n^3} \right) + 
\calO(\alphaeff^6) \,.
\end{align}
Here, $f(n,\varkappa)$ 
is the dimensionless Dirac energy~\cite{SwDr1991a,*SwDr1991b,*SwDr1991c}.

\end{widetext}

\begin{figure}[th!]
\centerline{\includegraphics[width=0.9\linewidth]{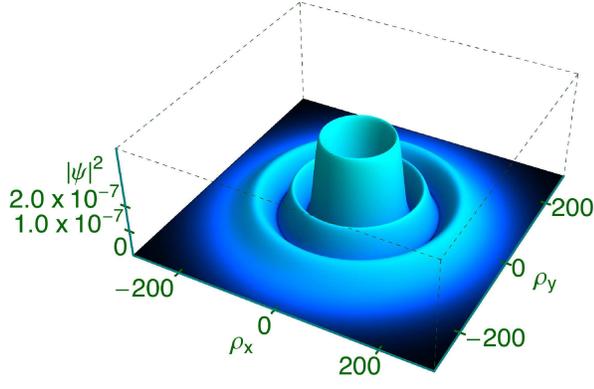}}
\caption{\label{fig2}
(Color online.)
Plot of the probability density $| \psi(\vec \rho) |^2$
for a bound electron in a state with
quantum numbers $n=12$, $\ell = 9$ and $m=9$,
in the field of the same
charged black hole as given in Fig.~\ref{fig1}.
The quantum numbers are $n = 12$,
$\ell = 9$ and $m = |\ell| = 9$,
and the polar angle is $\theta = 90^\circ$,
i.e., the plot pertains to the
($\rho_x,\rho_y)$ plane, with the dimensionless
vector-valued variable $\vec\rho$ 
being defined in Eq.~\eqref{scaling}.}
\end{figure}

We now consider a  numerical example.
Because  $Z_\Q$ and $Z_\q$ are the nuclear charge
numbers associated with the atoms, we
have $Z_\Q = +10$ and $Z_\q = -1$ for an orbiting
electron around a positively charged small black 
hole. 
%{\color{red} Small black holes are considered as possible 
% candidates for explanations of dark matter.}
Following Ref.~\cite{Je2014pra},
we thus consider a charged black hole with 
a mass $M$ equal to $10^{-13}$ times the mass of the earth,
\begin{equation}
M = 10^{-13} M_\oplus = 5.9742 \times 10^{11} \, {\rm kg}\,,
\end{equation}
and assume that $m = m_e$ (electron mass).
In the numerical calculations, we thus use the following coupling constants
\begin{subequations}
\begin{align}
\alphaeff =&\; 10 \, \alpha = 7.297\,352 \times 10^{-2} \,,
\\
\alpha_G =& \; 1.148\,884\times 10^{-3} \,,
\\
\alpha_Q =& \; 1.278\,353 \times 10^{-45} \,.
\end{align}
\end{subequations}
Here, in order to ensure the reproducibility of the results given 
below, we assume all decimal places given assumed to be exact,
even if the Newton's gravitational constant is currently known
only to one part in $10^4$ (see Table~XL of Ref.~\cite{MoTaNe2012}).
We consider two atomic states with quantum numbers
\begin{subequations}
\begin{align}
n_1 = 12 \,, \quad
\ell_1 = 9 \,, \quad
j_1 = \frac{19}{2} \,, \quad
\varkappa_1 = -10 \,,
\\
n_2 = 12 \,, \quad
\ell_2 = 9 \,, \quad
j_2 = \frac{17}{2} \,, \quad
\varkappa_2 = 9 \,,
\end{align}
\end{subequations}
%
% 1 - \frac{\alphaeff^2}{288} - \frac{\alphaeff^4}{92160}
The energy formula~\eqref{ERN} evaluates to 
\begin{align}
\label{E1}
E_1 =& \; m \left( 
\left( 1 + \frac{\alphaeff^2}{(2 + \sqrt{100 - \alphaeff^2})^2} \right)^{-1/2}
- \frac{\alpha_G \, \alphaeff}{144}
\right.
\nonumber\\
& \; \left.
- \frac{59 \alpha_G \, \alphaeff^3}{1\,313\,280}
+ \frac{\alpha_Q \, \alphaeff^2}{32\,832}
+ \frac{\alpha_Q \, \alphaeff^4}{3\,878\,280} \right) \,,
\end{align}
which translates into SI mksA units as follows
(the coefficient term which generates the contribution is 
given separately as a subscript of each item),
\begin{align}
\label{E1eV}
E_1 =& \; m_e \, c^2 +
\left(
\left. -9.44855 \right|_{\rm Dirac-Coulomb} 
- \left. 0.29751 \right|_{\alpha_G \, \alphaeff} 
\right.
\nonumber\\
& \; - \left. 1.02491 \times 10^{-5} \right|_{\alpha_G \, \alphaeff^3} 
\nonumber\\
& \; + \left. 1.05951 \times 10^{-46} \right|_{\alpha_Q \, \alphaeff^2} 
\nonumber\\
& \; 
\left.
+ \left. 4.77631 \times 10^{-51} \right|_{\alpha_Q \, \alphaeff^4} 
\right) \, {\rm eV} \,.
\end{align}
It becomes clear that the backaction effect of the space-time curvature
induced by the charged particle, parameterized by $\alpha_Q$, only is a small
effect in our example case, but still constitutes a conceptually 
interesting correction.
The energy of the second bound state 
considered for our example calculation is 
\begin{align}
\label{E2}
E_2 =& \; m \left(
\left( 1 + \frac{\alphaeff^2}{(3 + \sqrt{81 - \alphaeff^2})^2} \right)^{-1/2}
- \frac{\alpha_G \, \alphaeff}{144}
\right.
\nonumber\\
& \; \left.
- \frac{43 \alpha_G \, \alphaeff^3}{787\,968}
+ \frac{\alpha_Q \, \alphaeff^2}{32\,832}
+ \frac{23\,\alpha_Q \, \alphaeff^4}{66\,977\,280} \right) \,.
\end{align}
Here, the breakdown of contributions is as follows,
\begin{align}
\label{E2eV}
E_2 =& \; m_e \, c^2 + 
\left(
\left. -9.44860 \right|_{\rm Dirac-Coulomb}
- \left. 0.29751 \right|_{\alpha_G \, \alphaeff}
\right.
\nonumber\\
& \; - \left. 1.24495  \times 10^{-5} \right|_{\alpha_G \, \alphaeff^3}
\nonumber\\
& \; + \left. 1.05951 \times 10^{-46} \right|_{\alpha_Q \, \alphaeff^2}
\nonumber\\
& \;
\left.
+ \left. 6.36110 \times 10^{-51} \right|_{\alpha_Q \, \alphaeff^4}
\right) \, {\rm eV} \,.
\end{align}
The first terms on the right-hand sides of Eqs.~\eqref{E1eV}
and~\eqref{E2eV} contain the electron rest mass.
An expansion of the energies given in Eqs.~\eqref{E1} and~\eqref{E2}
in powers of $\alphaeff$ reveals that they differ only 
by the fine structure (i.e., at order $\alphaeff^4$ for terms 
free of $\alpha_G$ and $\alpha_Q$).

%
% Conclusions
%
\section{Conclusions}
\label{sec5}

In this paper, we find the nonrelativistic limit of the Dirac equation
coupled to a statically charged gravitational center.
To carry out this calculation we first have to derive the
Dirac--Reissner--Nordstr\"om Hamiltonian.
The derivation requires that we first transform the metric, and
consequently the potential, into isotropic coordinates, and then
couple the Dirac equation to both the curved space--time and the
electrostatic potential (see Sec.~\ref{sec2a} and Appendix~\ref{appa}).
The derivation of the Reissner--Nordstr\"om metric 
in isotropic coordinates could be of interest in a wider context.
Starting from generalized Dirac Hamiltonian, we find the nonrelativistic
limit by applying the Foldy--Wouthuysen program
(Secs.~\ref{sec2b} and~\ref{sec3}).
We carry out the transformation keeping terms up to the fourth order in 
$\alphaeff$, where $\alphaeff = -Z_\Q \, Z_q \, \alpha$ 
is an effective coupling constant for the 
bound system [see Eq.~\eqref{defalphaeff}].
% Then, $|\vec p|\sim \alphaeff m$ and $(1/r)\sim \alphaeff \, m$
% [see Eq.~\eqref{rpalpha}]. 
Furthermore, 
we keep terms of first order in the gravitational
constant $G$, i.e., first order in the effective 
coupling constants $\alpha_G$ and $\alpha_Q$ defined 
in Eqs.~\eqref{defalphaG} and~\eqref{defalphaQ}.

The Foldy--Wouthuysen transformation of the 
Dirac--Reissner--Nordstr\"{o}m Hamiltonian is 
carried out in Sec.~\ref{sec3} (see also Appendix~\ref{appb}),
and the structure of the resulting bound-state spectrum is 
analyzed in Sec.~\ref{sec4}.
The final result for the 
Foldy--Wouthuysen transformed Hamiltonian 
is given in Eq.~\eqref{HFWRN};
it contains a number of familiar terms.
As should be expected, when we remove the charge of the gravitational
center ($\alphaeff = \alpha_Q = 0$,
but $\alpha_G \neq 0$) we recover the nonrelativistic limit of
Dirac--Schwarzschild Hamiltonian~\cite{JeNo2013pra,Je2014pra}.
Additionally, if we were to neglect the gravitational terms 
($\alpha_G = \alpha_Q = 0$),
the nonrelativistic limit of the Dirac--Coulomb Hamiltonian is
recovered.
We also find terms additional terms which are 
due to the presence of the center charge which curves space--time,
as manifest in the differences of the Schwarzschild and the 
Reissner--Nordstr\"{o}m metric.
After the Foldy--Wouthuysen transformation, 
one recognizes these terms as mixing terms, 
proportional to a the product of a
gravitational coupling ($\alpha_G$ or $\alpha_Q$)
and a power of the effective electromagnetic coupling 
constant $\alphaeff$.

The result for the 
Dirac--Reissner--Nordstr\"{o}m Hamiltonian given in 
Eq.~\eqref{HFWRN} can naturally be written as the 
sum of three contributions, 
a free Hamiltonian $H_\F$ (with relativistic corrections),
a modified Dirac--Coulomb Hamiltonian $H'_{\rm DC}$, 
and a modified Dirac--Schwarzschild gravitational
potential term $H'_{\rm DS}$
[see Eqs.~\eqref{HDCp} and~\eqref{HDSp}].
There are perturbations to the Coulomb potential
due to the curvature of space--time, resulting from both the mass
and the charge of the gravitational center, 
i.e., proportional to both $r_\s$ as well as $r_\Q^2$
[see Eq.~\eqref{HDCp}].
The leading gravitational term 
(proportional to $r_\s$)
and the leading electro--gravitational mixing term 
(proportional to $r_\Q^2$) enter with 
opposite sign in both $H'_{\rm DC}$ 
and $H'_{\rm DS}$. 
$H'_{\rm DC}$ breaks the particle-antiparticle 
symmetry, while $H'_{\rm DS}$ conserves it.
The electric field corresponding 
to the Coulomb potential leads to a nonvanishing 
energy-momentum tensor, which modifies the 
space-time curvature around the charged black hole;
hence the differences of the Schwarzschild and 
Reissner--Nordstr\"{o}m metrics.
In turn, the metric enters the formulation of the 
generalized Dirac Hamiltonian, which 
naturally contains terms due to the modified 
space-time curvature, i.e., proportional to 
$r_\Q^2$. These higher-order (in $\alphaeff$) 
correction terms for bound states resulting from these
terms are clearly identified after a Foldy--Wouthuysen transformation.

The gravitational corrections to a Coulomb bound system,
which are derived here using a rigorous approach,
are of interest for a number of reasons.
Space-time noncommutativity is a concept inspired by string theory
(see Ref.~\cite{SeWi1999}), which could be of relevance
in the unification of gravity with the other fundamental
interactions.
As shown in Ref.~\cite{ChSJTu2001}, space-time noncommutativity
may ultimately induce certain shifts of energy levels in atomic
systems which may be detectable in the future.
Here, the leading fully relativistic gravitational corrections are
derived using a rigorous approach which does not
require space-time quantization.
Second, micro black holes have been proposed in
various contexts of physics, including
candidates for dark matter~\cite{GrBaAl2004,GrBa2008,DoEr2014}
and even classes of novel phenomena at
accelerators~\cite{JaBuWiSa2000}.
For charged micro black holes, a
quantum mechanical description requires the
use of the Reissner--Nordstr\"{o}m metric.

%
% Acknowledgements
%
\section*{Acknowledgements}

This research has been supported by a Missouri Research Board Grant.
Support from the National Science Foundation (Grant No. PHY–1403973) also is
gratefully acknowledged.

\appendix

%
% The Reissner--Nordstr\"om Metric
%
\section{The Reissner--Nordstr\"om Metric}
\label{appa}

The derivation of the Reissner--Nordstr\"om metric, which describes the
curvature of space--time due to a charged gravitational center, is generally
left as an exercise in textbooks (problem 6.3 in~\cite{Wa1984} for example). 
It is also possible to find unpublished works which detail the
derivation~\cite{Ma2009}. 
Here we briefly outline the derivation, and the assumptions made.

We are interested in a spherically symmetric, statically charged,
stationary black hole.  
Such a black hole will result in a static, spherically symmetric space--time. 
From Eq.~(6.1.5) of~\cite{Wa1984} we know that the
metric of such a space--time is of the form
\begin{align}
\dd s^2=&\,f(\rhosymb)\,\dd t^2 - h(\rhosymb)\,\dd\rhosymb^2 
- \rhosymb^2\,\dd\Omega\,,\\
\dd\Omega=&\,\dd\theta^2+\sin^2\theta\,\dd\varphi^2\,,
\end{align}
where we have adjusted from the ``East--coast'' convention used in
\cite{Wa1984} to the ``West-coast'' convention we use in this paper.
To derive the Reissner--Nordstr\"om metric we use the Einstein field equation
\begin{equation}
R_{\mu\nu}-\frac12g_{\mu\nu}R=8\pi G T_{\mu\nu}\,,
\end{equation}
which is equivalent to
\begin{equation}
\label{EinsteinEquation}
R_{\mu\nu}=8\pi G\left(T_{\mu\nu}-\frac12g_{\mu\nu}T\right)\,,
\end{equation}
where $R_{\mu\nu}$ is the Ricci tensor and $T_{\mu\nu}$ is the electromagnetic 
stress--energy tensor.
Using the ``West--coast'' convention, the electromagnetic stress--energy tensor
is
\begin{equation}
\label{StressEnergyTensor}
T_{\mu\nu}=-\left(F_{\mu\lambda}{F_\nu}^\lambda
-\frac14g_{\mu\nu}F^{\kappa\lambda}F_{\kappa\lambda}\right)\,.
\end{equation}
Notice that this equation has the opposite sign as compared to 
the stress--energy tensor
in the ``East--coast'' convention [see Eq.~(5.22) of~\cite{MiThWh1973}].
With this definition it is trivial to show that $T={T_\mu}^\mu=0$, and
Eq.~\eqref{EinsteinEquation} becomes
\begin{equation}
\label{RTrelation}
R_{\mu\nu}=8\pi G\,T_{\mu\nu}\,.
\end{equation}
We now need to calculate for the components of the Ricci tensor and the 
components of the electromagnetic stress--energy tensor.
By definition, the Ricci tensor is
\begin{equation}
R_{\mu\nu}=R^\lambda_{\mu\lambda\nu}
=\partial_\lambda\Gamma^\lambda_{\mu\nu}
-\partial_\nu\Gamma^\lambda_{\mu\lambda}
+\Gamma^\lambda_{\sigma\lambda}\Gamma^\sigma_{\mu\nu}
-\Gamma^\lambda_{\sigma\nu}\Gamma^\sigma_{\mu\lambda}\,,
\end{equation}
where $\Gamma^\lambda_{\mu\nu}$ is the Christoffel symbol.
The non--zero components of the Ricci tensor are then found to be
\begin{subequations}
\begin{align}
R_{tt}=&\,\frac{f'}{{\rhosymb}\,h}-\frac{f'\,^2}{4\,fh}
-\frac{f'\,h'}{4\,h^2}+\frac{f''}{2\,h}\,,\\
R_{\rhosymb\rhosymb}=&\,\frac{f'\,^2}{4\,f^2}+\frac{h'}{\rhosymb\,h}
+\frac{f'\,h'}{4\,fh}-\frac{f''}{2\,f}\,,\\
\label{R_thetatheta}
R_{\theta\theta}=&\,1-\frac1h-\frac{\rhosymb\,f'}{2\,fh}
+\frac{\rhosymb\,h'}{2\,h^2}\,,
\quad R_{\phi\phi}=R_{\theta\theta}\sin^2\theta\,,
\end{align}
\end{subequations}
where the prime indicates differentiation with respect to $\rhosymb$.

We now consider the electromagnetic stress--energy tensor.
As we are considering a static spherically symmetric electric field 
(without currents or magnetic fields), 
the only non--zero components of the field strength tensor are
\begin{equation}
F_{t\rhosymb}=-F_{\rhosymb t}=E(\rhosymb)\,.
\end{equation}
The electric field outside of a uniformly charged 
sphere is given as
\begin{equation}
E(\rhosymb)=\frac{Q}{4\pi\,\rhosymb^2}\,.
\end{equation}
Using Eq.~\eqref{StressEnergyTensor} we can now calculate the components of
the electromagnetic stress--energy tensor.
The non--zero components are
\begin{subequations}
\begin{align}
T_{tt}=&\,\frac1h\frac{Q^2}{32\pi^2\,\rhosymb^4}\,,\\
T_{\rhosymb\rhosymb}=&\,-\frac1f\frac{Q^2}{32\pi^2\,\rhosymb^4}\,,\\
\label{T_thetatheta}
T_{\theta\theta}=&\,\frac1{f\,h}\frac{Q^2}{32\pi^2\,\rhosymb^2}\,,\quad
T_{\phi\phi}=T_{\theta\theta}\sin^2\theta\,.
\end{align}
\end{subequations}
Notice that $f^{-1}T_{tt}+h^{-1}T_{\rhosymb\rhosymb}=0$, therefore
\begin{equation}
\frac1fR_{tt}+\frac1hR_{\rhosymb\rhosymb}=\frac1{r\,fh^2}(fh)'=0\,,
\end{equation}
from which we conclude that
\begin{equation}
f=K\,h^{-1}\,,
\end{equation}
where $K$ is a constant.
As was done in Sec.~6.1 of~\cite{Wa1984}, we can gauge away $K$ by re--scaling
the time coordinate as $dt\to\sqrt{K}dt$.
Thus $h=f^{-1}$, and Eqs.~\eqref{R_thetatheta} and~\eqref{T_thetatheta}
become
\begin{align}
R_{\theta\theta}=1-\partial_\rhosymb\left(\rhosymb\,f\right)\,,\quad
T_{\theta\theta}=\frac{Q^2}{32\pi^2\,\rhosymb^2}\,,
\end{align}
respectively.
Applying these equations to Eq.~\eqref{RTrelation} we find
\begin{equation}
1-\partial_\rhosymb\left(\rhosymb\,f\right)=\frac{r_\Q^2}{\rhosymb^2}\,,
\qquad
r_{\rm Q}^2=\frac{G\,Q^2}{4\pi}\,.
\end{equation}
This equation is solved by
\begin{equation}
f=1+\frac{C}\rhosymb+\frac{r_\Q^2}{\rhosymb^2} \,.
\end{equation}
If we set $Q=0$ then we should recover the Schwarzschild metric.
Thus $C=-r_{\rm s}=2GM$, and the Reissner--Nordstr\"om metric is found to be
\begin{align}
\dd s^2=&\,\left(1-\frac{r_\s}{\rhosymb}+\frac{r_\Q^2}{\rhosymb^2}\right)\dd t^2
\nonumber\\&\,
-\left(1-\frac{r_\s}{\rhosymb}+\frac{r_\Q^2}{\rhosymb^2}\right)^{-1}\dd\rhosymb^2
-\rhosymb^2\,\dd\Omega^2\,.
\end{align}
The simplified approach to the derivation of the 
metric taken here, makes extensive use of the known 
solution for the Schwarzschild geometry and 
leads to a streamlined derivation.

%
% Derivation of the Hamiltonian
%
\section{Derivation of the Hamiltonian}
\label{appb}

Here we follow the notation utilized in
Refs.~\cite{JeNo2013pra,NoJe2015tach}.
The flat--space--time Dirac gamma matrices are
denoted with a tilde ($\fs$) while the curved--space--time Dirac gamma
matrices are written with an overline ($\cs$).
The notation for the curved-space--time matrices is inspired 
by the covariant structure of their anticommutator,
expressed in Eq.~\eqref{covgamma},
and the tensor (``vector'') structure 
is denoted by the overline.
By contrast, from the point of view of general
relativity, the flat-space matrices 
can be regarded as ``modified'' $\gamma$ matrices,
hence the tilde. The flat-space matrices are 
used in the Dirac representation,
\begin{subequations}
\label{dirac_rep}
\begin{align}
\wtilde\gamma^0 =& \;
\left( \begin{array}{cc} \mathbbm{1}_{2\times2} & 0 \\
0 & -\mathbbm{1}_{2\times2} \end{array} \right) \,,
\quad
\wtilde\gamma^1 =
\left( \begin{array}{cc} 0 & \sigma^1 \\ -\sigma^1 & 0  \end{array} \right) \,,
\\[0.007ex]
\wtilde\gamma^2 = & \;
\left( \begin{array}{cc} 0 & \sigma^2 \\ -\sigma^2 & 0  \end{array} \right) \,,
\qquad
\wtilde\gamma^3 =
\left( \begin{array}{cc} 0 & \sigma^3 \\ -\sigma^3 & 0  \end{array} \right) \,,
\\[0.007ex]
\wtilde\gamma^5 =& \;
\ii \, \wtilde\gamma^0 \, \wtilde\gamma^1 \,
\wtilde\gamma^2 \, \wtilde\gamma^3
= \left( \begin{array}{cc} 0 & \mathbbm{1}_{2\times2} \\
\mathbbm{1}_{2\times2} & 0 \end{array} \right) \,,
\end{align}
\end{subequations}
where the $\sigma^i$ are the $(2 \times 2)$ Pauli matrices~\cite{BjDr1964}.

As in Appendix C of Ref.~\cite{NoJe2015tach} 
we draw inspiration from Ref.~\cite{Bo2011}
and use the capital Latin indices $A,B,C,...=0,1,2,3$ to denote the
local Lorentz frame (``anholonomic basis'') and $I,J,K,...=1,2,3$ for spatial
coordinates in the anholonomic basis.
The Greek indices $\mu,\nu,\lambda,...=0,1,2,3$ denote the global coordinates,
while the lower case Latin indices $i,j,k,...=1,2,3$ are used for the global
spatial coordinates.
As in Refs.~\cite{JeNo2013pra,NoJe2015tach} 
we use the ``West Coast'' convention for the flat--space--time
metric, denoted as ${\tilde g}_{AB} = 
\eta_{AB}=\eta^{AB}={\rm diag}(1,-1,-1,-1)$.
The curved--space--time metric is denoted using $\overline g_{\mu\nu} = 
g_{\mu\nu}$, without the need
for an overline, as $\eta$ denotes the flat--space--time metric.
From Eq.~\eqref{isoRNmetric} we know that the curved--space--time
metric is
\begin{subequations}
\label{csmetric}
\begin{align}
g_{\mu\nu}=&\,{\rm diag}\left(w^2,-v^2,-v^2,-v^2\right)\,,\\
g^{\mu\nu}=&\,{\rm diag}\left(w^{-2},-v^{-2},-v^{-2},-v^{-2}\right)\,.
\end{align}
\end{subequations}
This metric has the same structure as the isotropic Schwarzschild metric in the
Eddington reparameterization [see Ref.~\cite{Ed1924} 
and Eq.~(8) of Ref.~\cite{JeNo2013pra},
as well as Eqs.~(C6) and (C7) of Ref.~\cite{NoJe2015tach}].
However,
the functions $w=w(r)$ and $v=v(r)$ are different
for the Reissner--Nordstr\"{o}m geometry.
The curved--space--time Dirac gamma matrices can be expressed in terms
of the flat--space--time Dirac gamma matrices as
\begin{align}
\cs^\mu(x)=e^\mu_A(x)\fs^A\,,\quad
\cs_\mu(x)=e^A_\mu(x)\fs_A\,,
\end{align}
where $e^\mu_A(x)$ are the vierbein coefficients.
By definition, the curved--space--time Dirac gamma matrices must satisfy the
condition
\begin{align}
\label{covgamma}
\left\{\cs_\mu(x),\cs_\nu(x)\right\}=&\,2g_{\mu\nu}\,,
\end{align}
from which we find
\begin{subequations}
\begin{align}
g_{\mu\nu}=&\,\frac12\left\{\cs_\mu(x),\cs_\nu(x)\right\}
=e^A_\mu(x)\,e^B_\mu(x)\,\eta_{AB}\,,\\
g^{\mu\nu}=&\,\frac12\left\{\cs^\mu(x),\cs^\nu(x)\right\}
=e_A^\mu(x)\,e_B^\mu(x)\,\eta^{AB}\,.
\end{align}
\end{subequations}
The vierbein coefficients that satisfy these equations are
\begin{subequations}
\begin{align}
e^0_\mu=&\,\delta^0_\mu\,w\,,\quad
e^I_\mu=\delta^I_\mu\,v\,,\\
e^\mu_0=&\,\delta^\mu_0\,\frac1w\,,\quad
e^\mu_I=\delta^\mu_I\,\frac1v\,.
\end{align}
\end{subequations}
Here $\delta^\mu_A$ and $\delta^A_\mu$ denote the Kronecker delta.

As is well known, when formulating the Dirac equation
in curved--space--time, one replaces the $\fs^A\to\cs^\mu$ and
$\partial_A\to\nabla_\mu$ 
(see Refs.~\cite{SiTe2005,Si2008pra,ObSiTe2009,ObSiTe2011,ZaMB2012,Je2013,JeNo2013pra,%
JeNo2014jpa,NoJe2015tach}), where
\begin{align}
\label{CDspinor}
\nabla_\mu=&\,\partial_\mu-\Gamma_\mu\,,\\
\Gamma_\mu=&\,\frac\ii4\omega^{AB}_\mu\,\sigma_{AB}\,,\quad
\sigma_{AB}=\frac\ii2\left[\fs_A,\fs_B\right]\,,\\
\label{RicciRotCoef}
\omega^{AB}_\nu=&\,e^A_\mu\nabla_\nu\,e^{\mu B}
=e^A_\mu\partial_\nu\,e^{\mu B}+e^A_\mu\Gamma^\mu_{\nu\lambda}\,e^{\lambda B}\,.
\end{align}
For absolute clarity, we emphasize that 
the covariant derivative ``$\nabla$''
in Eq.~\eqref{CDspinor} acts on a spinor,
while the ``$\nabla$'' in Eq.~\eqref{RicciRotCoef} is the holonomic 
covariant derivative acting on a vector, defined as 
$\nabla_\nu A^\mu=\partial_\nu A^\mu+\Gamma^\mu_{\nu\lambda}A^\lambda$.
Furthermore, $\Gamma_\mu$ is the affine spin--connection matrix,
while $\Gamma^\mu_{\nu\lambda}$ are the Christoffel symbols and
$\omega^{AB}_\nu$ are the Ricci rotation coefficients.
This transformation ensures that the curved--space--time Dirac equation
is invariant under a Lorentz transformation.

When coupling the electrostatic potential to the Dirac equation in 
flat--space--time one replaces $\ii\partial_B\to\ii\partial_B-q\,A_B$,
where $q$ is the charge of the particle~\cite{BjDr1964}.
This is easily generalized to curved--space--time as
$\ii\nabla_\mu\to\ii\nabla_\mu-q\,A_\mu$.
Thus the Dirac equation in curved space--time, coupled to an electrostatic 
potential, is
\begin{equation}
\left(\cs^\mu\left(\ii\nabla_\mu-q\,A_\mu\right)-m\right)\psi=0\,.
\end{equation}
We know that the only non--zero term of the electrostatic potential is $A_0$
(Eq.~\eqref{IsoPot}), thus we can somewhat simplify the Dirac equation to
\begin{equation}
\left(\ii\cs^\mu\partial_\mu-\ii\cs^\mu\Gamma_\mu-\cs^0q\,A_0
-m\right)\psi=0\,.
\end{equation}
Multiplying by $\cs^0$ on the left, and rearranging the equation we obtain
\begin{align}
\label{intham}
\ii&\,\left(\cs^0\right)^2\partial_0\psi\nonumber\\
&= \left(-\ii\,\cs^0\cs^i\partial_i 
+ \ii\,\cs^0\cs^\mu\Gamma_\mu
+ \left(\cs^0\right)^2 q\,A_0
+ \cs^0m\right)\psi\,.
\end{align}
We now utilize our vierbein, along with an explicit calculation of the 
affine spin--connection matrix to find
\begin{subequations}
\label{impident}
\begin{align}
\cs^0=&\,\frac1w\fs^0\,,\quad
\left(\cs^0\right)^2=\frac1{w^2}\,,\quad
\cs^0\cs^i=\frac1{vw}\fs^i\,,\\
\cs^0\cs^\mu\Gamma_\mu
=&\,-\frac{\vec\alpha\cdot\vec\nabla\,w}{2vw^2}
-\frac{\vec\alpha\cdot\vec\nabla\,v}{v^2w}\,.
\end{align}
\end{subequations}
The form of these results are in full agreement with 
Refs.~\cite{JeNo2013pra,NoJe2015tach},
the only differences coming from the definitions of the functions
$w$ and $v$.
Here $\vec\alpha=\fs^0\vec\fs$.
Applying the results found in Eq.~\eqref{impident} to Eq.~\eqref{intham}
and multiplying by $w^2$ on the left we find $\ii\,\partial_t\,\psi=H\,\psi$
where
\begin{align}
H=\frac wv\adp+\frac{\vec\alpha\cdot\left[\vec p,w\right]}{2v}
+\frac wv\frac{\vec\alpha\cdot\left[\vec p,v\right]}{v}
+q\,A_0+\beta\,m\,w\,,
\end{align}
and $\beta=\fs^0$.
As done in \cite{JeNo2013pra}, we rescale the spatial coordinates according to
$\psi'=v^{3/2}\psi$, and $H'=v^{3/2}H\,v^{-3/2}$, to find the Hermitian
Hamiltonian
\begin{equation}
H'=\frac12\left\{\adp,\frac wv\right\}+q\,A_0+\beta\,m\,w\,.
\end{equation}
Finally we use our approximations from Eq.~\eqref{approxwv} to find
\begin{equation}
\frac wv\approx1-\rr+ \frac{3 r_\Q^2}{4 r^2}\,,
\end{equation}
and apply them, along with Eq.~\eqref{IsoPot} to the Hamiltonian to find
the Dirac--Reissner--Nordstr\"om Hamiltonian, to the first order
in $G$, as
\begin{align}
H_\RN =&\,\frac12\left\{\adp,\left(1-\rr+
\frac{3 r_\Q^2}{4 r^2}\right)\right\}
\nonumber\\&\,
+\frac{Z_\Q\, Z_\q\,\alpha}r\left(1-\rrt+\rcf\right)
\nonumber\\&\,
+\beta\,m\left(1-\rrt+\rct\right)\,,
\end{align}
where we use $q\,Q=4\pi\,Z_\Q\,Z_\q\,\alpha$.
Here, $Z_\Q$ is the nuclear charge number of the central 
gravitational object (charge $Q$), while
$Z_\q$ is the nuclear charge number associated with 
the test charge $q$,
and $\alpha$ is the fine--structure constant.

\end{document}